\documentstyle[aps]{revtex}

\begin{document}
\title{The Algebraic Model for scattering in three-s-cluster systems.\\
II. Resonances in the three-cluster continuum of $^{6}$He and $^{6}$Be}
\author{V. Vasilevsky, A. V. Nesterov, F. Arickx, J. Broeckhove}
\address{Bogolyubov Institute for Theoretical Physics, Kiev, Ukraine\\
Universiteit Antwerpen (RUCA), Dept. of Mathematics and Computer Science,\\
Antwerp, Belgium}
\date{\today}
\maketitle

\begin{abstract}
The resonance states embedded in the three-cluster continuum of $^{6}$He and 
$^{6}$Be are obtained in the Algebraic Version of the Resonating Group
Method. The model accounts for a correct treatment of the Pauli principle.
It also provides the correct three-cluster continuum boundary conditions by
using a Hyperspherical Harmonics basis. The model reproduces the observed
resonances well and achieves good agreement with other models. A better
understanding for the process of formation and decay of the resonance states
in six-nucleon systems is obtained.
\end{abstract}

\section{Introduction}

In this paper we investigate the three-cluster continuum of nuclei $^{6}$He
and $^{6}$Be, determined by the three-cluster configurations $\alpha +n+n$
and $\alpha +p+p$. This is a first calculation of the Algebraic Model for
scattering in three-s-cluster systems, as described in \cite{kn:ITP+RUCA1}.
Our objective is to highlight the quantitative and qualitative
characteristics of Algebraic Model three-cluster calculations, as well as to
produce accurate results for the astrophysically relevant resonances in $%
^{6} $He and $^{6}$Be.

As the Algebraic Model for three-cluster scattering leads to a set of
coupled-channel Schr\"{o}dinger equations, we will transform the $S$-matrix
to a diagonal form. This is usually referred to as the eigenchannel
representation of the $S$-matrix \cite{kn:ITP+RUCA1}. We will derive the
position and width of the resonances in the usual way i.e. by scanning the
corresponding eigenphase-shifts $\delta $ as a function of energy for
resonance behavior. To be precise, we determine position and widths through
the conditions 
\begin{equation}
\frac{d^{2}\delta }{dE^{2}}|_{E=E_{r}}=0,\quad \Gamma =2\left[ \frac{d\delta 
}{dE}\right] ^{-1}|_{E=E_{r}}  \label{eq:ResConds}
\end{equation}

Alternative methods have been suggested, such as the Complex Scaling Method 
\cite{kn:CSM-ABC1,kn:CSM-ABC2} (CSM). This method is based on a complex
scaling of the Hamiltonian which reduces the scattering problem to an
eigenvalue problem of a complex non-Hermitian Hamiltonian. The method was
shown to reproduce efficiently both very narrow as broad resonance states.
The drawback to the method is connected to the oscillatory behavior of the
resonance. To get a stable and correct result for the position of the
resonance, this behavior should be properly reproduced, requiring the
calculation of huge complex Hamiltonian matrices. An interesting review,
covering CSM as well as other methods for the determination of resonance
characteristics, is presented in \cite{kn:CSM-Ho}.

A third method of interest for evaluating the poles of the $S$-matrix called
the Continuation in the Coupling Constant (CCCM) and is appropriate for
resonances near to the threshold \cite{kn:kukulin+89}. A coupling parameter $%
\lambda $ is introduced such that for $\lambda =0$ repulsion in the system
is suppressed, and for $\lambda =1$ the original system is recovered. The
parameter $\lambda $\ is varied from zero up to a critical value $\lambda
_{c}$\ at which value the bound state crosses the threshold; by analytic
continuation up to $\lambda =1$\ one then obtains a good approximation for a
pole of the S-matrix in the complex energy plane.

In our work the nature and properties of the low-lying 0$^{+}$-and 2$^{+}$%
-resonances in $^{6}$Be and the 2$^{+}$-resonance state in $^{6}$He will be
studied in detail. The results of this paper are the next step in an
investigation of six-nucleon systems in the three-cluster model, initiated
in \cite{kn:Vasil96,kn:Vasil97}. In \cite{kn:Vasil96,kn:Vasil97} we focused
mainly on the\ problems of convergence and of the selection of the dominant
part of the three-cluster subspace for the bound state properties in $^{6}$%
He and $^{6}$Li. In the current paper we consider an appropriate mixture of
basis functions tailored to describe both the internal and asymptotic
regions. In particular, as the asymptotic behavior is reproduced most
naturally by a Hyperspherical Harmonics (HH) description, the basis has to
account for a sufficiently large set of Hyperspherical Harmonic wave
functions. As already indicated above, the resonance position and widths
will be obtained through the standard eigenphase analysis.

Several models and methods have already been applied to investigate the
resonance states of $^{6}$He and $^{6}$Be. In the series of papers \cite
{kn:Zhuk93,kn:Dani91,kn:Dani97}, these systems were considered as
three-particle ones, and consequently antisymmetrization was neglected.
These authors used an effective interaction between the alpha-particle and a
nucleon, and a NN realistic potential between the two nucleons. In these
papers the Hyperspherical Harmonics Method was used to describe the bound as
well as the three-particle continuum states. In \cite{kn:Csoto94} the CSM
was used to calculate the characteristic parameters of the resonance states
in $^{6}$He and $^{6}$Be. This was done within a three-cluster model, in
which full antisymmetrization was taken into account, and the interactions
between clusters was obtained from the Minnesota NN-potential. In \cite
{kn:CSM-tanaka} it was demonstrated that the method of continuation in the
coupling constant (CCCM), used for the $\alpha +N+N$ configurations in $^{6}$%
He and $^{6}$Li, reproduces results, which are very close to those obtained
by the CSM. We will compare our results to all three alternative methods.

\section{Details of the calculation}

To model the interaction between nucleons the Volkov potential \cite
{kn:Volk65} has been used throughout all calculations. As was shown in \cite
{kn:Vasil96,kn:Vasil97} it provides an acceptable description for $^{6}He$
within the three-cluster model. The effective Volkov potential consists of
central forces only, so that total angular momentum $L$, total spin $S$ and
total isospin $T$ are good quantum numbers. We will therefore consider only
three-cluster configurations $\alpha +N+N$ with $S=0$ and $T=1$. This is
known to be the most prominent and dynamically distinguished spin-isospin
state for the resonances of interest in the nuclei considered in this work.
The Coulomb interaction has also been included, as it is to a great extent
responsible for reproducing the 0$^{+}$ resonance state in $^{6} $Be.

The oscillator radius $b$, associated with the basis functions for the
nuclear state, is the only free parameter for the Algebraic Model
calculation. We have chosen to fix $b$ by optimizing the ground-state energy
of the $\alpha $-particle with respect to the Volkov interaction. This leads
to a value of $b=1.37$ fm.

As explained in \cite{kn:ITP+RUCA1}, we have a choice of Jacobi coordinate
systems. This provides an additional degree of freedom that can be exploited
to simplify the calculations. For a cluster configuration $\alpha +N+N$ one
can consider the two Jacobi configurations displayed in Fig. \ref
{fig:figure1}. The first (referred to as the ``4+2''-configuration) is the\
most appropriate for the current calculation. Selection rules effect a
significant reduction in the number of basis functions. Indeed, quantum
numbers $S=0,T=1$ coincide for the full six-nucleon system and the
two-nucleon subsystem ($N+N$). Hence the relative motion wave function must
be an even function in the coordinate $q_{1}$ of this Jacobi system. This
means that only even angular momenta $l_{1}$ have to be considered.
Moreover, for positive parity states only even values of $l_{2}$, and for
negative parity states only odd values of $l_{2}$ should be taken into
account. With the second Jacobi configuration from Fig. \ref{fig:figure1}\
(the ``5+1'' one), these constraints are hard to meet and the full basis of
oscillator states would have to be considered.

Table \ref{tab:HH state} shows the number of HH's ($N_{h}$) of given
hypermomentum $K$ for $L^{\pi }=0^{+}$ and $L^{\pi }=2^{+}$ in the ``4+2''
Jacobi configuration. This table also indicates the total number of HH's so
far ($N_{c}$) with $K=K_{\min },$ $K_{\min }+2,\ldots ,K_{\max }$, i.e. it
shows the number of channels for a given $K_{\max }=K$.

\subsection{Overlap matrix elements}

As full antisymmetrization between all nucleons is taken into account within
the interaction region of the clusters, we first\ investigate the role of
the Pauli principle for three-cluster configurations. This is of importance
amongst others for marking the boundary between the internal (interaction)
region and the asymptotic region. No particle exchanges should occur in the
asymptotic region. In the Algebraic Model the antisymmetrization effects can
be visualized through the overlap and potential matrix elements. In
Resonating Group Method descriptions such as the integro-differential or the
Generator coordinate versions, overlap and potential matrix elements are
nonlocal functions of the coordinates, and are difficult to analyze. We use
the notations of \cite{kn:ITP+RUCA1} for the overlap matrix elements, and in
particular the shorthand notation $\nu _{0}$\ is used for the set $%
(l_{1}l_{2})LM$\ of quantum numbers: 
\begin{equation}
\left\langle n,\left( K,\nu _{0}\right) \left| \widehat{{\cal A}}\right|
n^{\prime },\left( K^{\prime },\nu _{0}\right) \right\rangle
\label{eq:OvlapMatrixElem}
\end{equation}
Non-zero matrix elements (\ref{eq:OvlapMatrixElem}) can be obtained from
states within the same many-particle oscillator shell only. As the
oscillator shells in the Hyperspherical description are characterized by $%
N=2n+K$ , the selection rule becomes $2n+K=2n^{\prime }+K^{\prime }$.

In Fig. \ref{fig:figure2}\ overlap matrix elements diagonal in $n$ for $L=0$
and hypermomenta $K=0$ and $K=2$ are shown for the six-nucleon three-cluster
system. One notices from this figure that the Pauli principle involves
oscillator states of at least the 25 lowest shells, for both diagonal and
off-diagonal matrix elements in $K$. The antisymmetrization effects are
visible in the deviation from unity for the diagonal matrix elements, and
the deviation from zero of the off-diagonal (in $K$) elements. These effects
decrease monotonically with higher $n$.

In Fig. \ref{fig:figure3} we compare matrix elements diagonal in $n$ and $K$ 
$\left\langle n,\left( K,\sigma ,L=0\right) \left| \widehat{{\cal A}}\right|
n,\left( K,\sigma ,L=0\right) \right\rangle $ for some of the $K$-values, $%
\sigma $ being a multiplicity quantum number for states with identical $K$.
Only those matrix elements where the Pauli principle effect is most
prominent have been shown. Some states with $K=4$ and $K=8$ are affected
more strongly by antisymmetrization than others. To understand this we note
that the Hyperspherical angles (corresponding to the hyperangular quantum
numbers $K,l_{1},l_{2},LM$) define the most probable triangular shape and
orientation in space of the three-cluster system. The HH's with $K=4,\sigma
=2$ (characterized by $l_{1}=l_{2}=2$) and $K=8,\sigma =3$ (characterized by 
$l_{1}=l_{2}=4$) seem to describe a triangular shape where one of the
nucleons is very close to the $\alpha $-particle.

For larger $K$-values the probability to find all clusters close to one
another within a hypersphere of fixed radius $\rho $ decreases, and one can
expect that HH's with large values of $K$ will play a diminishing role in
the calculations.

It interesting to compare overlap matrix elements for the three-cluster
configuration $\alpha +N+N$ with those of the two-cluster configurations $%
\alpha +2N$ (such as $\alpha +d$ in $^{6}Li$ or $\alpha +2n$ in $^{6}He$).
This comparison is shown in Fig. \ref{fig:figure4}, and it indicates that
the Pauli principle has a much larger ``range'' in the three-cluster than in
the two-cluster configuration.

\subsection{Potential matrix elements}

To investigate the Pauli effect on the potential energy of the three-cluster
configuration $\alpha +N+N$ we compare the potential matrix elements with
hypermomentum $K=0$ with full antisymmetrization against those in the
folding approximation, where antisymmetrization between clusters is
neglected. Figure \ref{fig:figure5} shows the diagonal potential matrix
elements $<n,\left( K=0,\nu _{0}\right) |V|n,\left( K=0,\nu _{0}\right) >$
and Fig. \ref{fig:figure6} shows the matrix elements along a fixed row $%
<n=50,\left( K=0,\nu _{0}\right) |V|n,\left( K=0,\nu _{0}\right) >$,
calculated both with and without antisymmetrization between clusters. One
notes that the folding model results are very close to the fully
antisymmetrized ones, especially for larger $n$.

In Fig. \ref{fig:figure7}\ we also display the potential matrix elements
between states of the two lowest values of hypermomentum $K=0$ and $K=2$.
One sees that the $K=2$ contribution is the largest. The potential energy
for $K=0$ is relatively small, and so is the coupling between $K=0$ and $K=2$
states.

The main conclusion is that, in the asymptotic region, the ``exact''
potential energy can be substituted with the folding approximation. It leads
to a considerable reduction in computational effort. We are led to the
following setup for three--cluster calculations. In the internal region,
consisting of states of the lower oscillator shells and with a large
probability to find the clusters close to one another, the fully
antisymmetrized potential energy is used. In the asymptotic region, where
the average distance between clusters is large, we neglect
antisymmetrization and use the folding model potential.

The folding model provides additional insight in the structure of the
interaction matrix. It is well known that neither the $\alpha +n$ nor the $%
n+n$ interaction can create a bound state in the corresponding subsystems of 
$^{6}He$. Only a full three--cluster configuration $\alpha +n+n$ contains
the necessary conditions to create a bound state. Within the folding model
the total potential energy is naturally split up in the individual
contributions from the three interacting pairs: the $\alpha $--particle and
the first neutron, the $\alpha $--particle and the second neutron and the
neutron-neutron pair. The folding model thus allows for a comparison of the
various contributions to the potential energy. For $K=0$ the first two
contributions are identical, and we only have to consider the $\alpha +n$
and the $n+n$ pairs. Figure \ref{fig:figure8} shows the diagonal matrix
element contribution of both components, and one notices that $\alpha +n$
represents the main contribution.

\subsection{The effective charge.}

When Coulomb forces are taken into account, one needs to determine the
effective charge in order to properly solve the Algebraic Model equations.
The effective charge unambiguously defines the effective Coulomb interaction
in each channel as well as the coupling between different channels. The
effective charge also defines the solutions in asymptotic region through
Sommerfeld parameter (see \cite{kn:ITP+RUCA1}). Using the approach suggested
in \cite{kn:ITP+RUCA1} the effective charges for the $0^{+}$- and $2^{+}$%
-states of $^{6}$Be were calculated. Part of the corresponding matrixes of $%
\left\| Z_{K}^{K^{\prime }}\right\| $ are displayed in tables \ref
{tab:table1} and \ref{tab:table2} respectively. One notices that the
diagonal matrix elements are much larger than the off-diagonal ones. This
justifies the approximation \cite{kn:ITP+RUCA1}\ of disregarding the
coupling of the channels in the asymptotic region.

It is interesting to compare the three-cluster effective charge\ with the
effective charge in the two-cluster configuration. For the latter we can
write 
\[
Z=Z_{1}Z_{2}e^{2}\sqrt{\frac{A_{1}A_{2}}{A_{1}+A_{2}}} 
\]
where $A_{1}$ and $Z_{1}$\ ($A_{2}$ and $Z_{2}$) are the respective mass and
charge of both clusters. For the configuration $\alpha +2p$ in $^{6}Be$, we
then obtain an effective charge 
\[
Z=\frac{8}{\sqrt{3}}e^{2}\simeq 6.65 
\]
which is independent on the angular momentum of the system. One notice that
the two-cluster effective charge is close to the diagonal matrix elements of 
$\left\| Z_{K}^{K^{\prime }}\right\| $. We can assume that, if\ in one of
the three-cluster channels the effective charge is very close to the
two-cluster one, it could indicate that the two protons move as an aggregate
in the asymptotic region. For the $2^{+}$-state we observe at least one
channel with this property, carrying the labels $K=4$, $l_{1}=2,$ $l_{2}=0$.

\section{Results}

\subsection{Definition of the model space}

The model space for the current calculations is primarily determined by the
total number of HH's in the internal and external region, and the number of
oscillator states. Different sets of HH's can be used in the internal and
asymptotic regions. An extensive set of HH's in the internal region will
provide a well correlated description of the three-cluster system due to the
coupling between states with different hypermomentum. The HH's in the
asymptotic region, which are exactly (without Coulomb) or nearly exactly
(Coulomb included) decoupled, are responsible for the richness in decay
possibilities.

In the current paper we restricted both internal and asymptotic HH sets
maximal hypermomentum value $K_{\max }^{(i)}=K_{\max }^{(a)}=10$. By
extending the respective subspaces up to these maximal values, we obtain a
fair indication of convergence.

We fixed the matching point between internal and asymptotic region at $N=50$%
. It is based on the observations of the previous sections, i.e. (1) it is
sufficiently large to allow the Pauli principle its full impact, and (2) it
is such that only possibly long-range effects of the potential are still
remaining in the asymptotic region, which can then be taken properly into
consideration in the defining three-term recursion formula for the
asymptotic expansion coefficients of the Algebraic Model \cite{kn:ITP+RUCA1}%
. The choice of matching point thus represents a proper compromise between
kinematical (i.e. Pauli) and dynamical (i.e. interaction) effects.

The resonance parameters are obtained from the eigenphase shifts obtained
from the eigenchannel representation (diagonalized form) of the $S$-matrix.

\subsection{Convergence study}

We first consider the influence of the number of asymptotic channels. We do
this by calculating the position and width of the 0$^{+}$-state in $^{6}$Be
using successively larger asymptotic subspaces. The first calculation
includes all HH's up to $K_{\max }^{(i)}=8$ and extends the asymptotic
subspace from $K_{\max }^{(a)}=0$ to $K_{\max }^{(a)}=8$. The second
calculation includes all HH's up to $K_{\max }^{(i)}=10$ and extends the
asymptotic subspace from $K_{\max }^{(a)}=0$ to $K_{\max }^{(a)}=10$. The
corresponding results are shown in tables \ref{tab:Conv_vs_asy_8} and \ref
{tab:Conv_vs_asy_10}. One learns from these results that a sufficient rate
of convergence has been obtained. Figure \ref{fig:figure9} displays the
first eigenphase shift as a function of energy for $K_{\max }^{(i)}=10$ and
a choice of $K_{\max }^{(a)}$ values from which results in the tables are
derived.

Whereas the inclusion of higher hypermomenta in the asymptotics shows a fast
and monotonic convergence, it is also clear from these results that a
sufficient number of HH's has to be used for a correct description of the
correlations in the internal state. To support this conclusion, we performed
a calculation, again for the 0$^{+}$-state in $^{6}$Be, in which only one HH
with value $K^{(a)}=0$ was used. In the internal region the number of HH's
was varied from $K^{(i)}=0$ up to $K^{(i)}=10$. These results appear in
table \ref{tab:Conv_vs_int} and corroborate the previous conclusion. In
particular they indicate that the effective potential obtained with the $%
K^{(i)}=0$ HH only, is unable to produce a resonance. Only after including a 
$K=2$ HH does the resonance appear, with an overestimated energy and width.
Further inclusion of higher HH states then lead to an acceptable convergence
in a monotonically decreasing fashion for both position and width of the
resonance.

\subsection{Comparisons}

In Fig. \ref{fig:figure10} the shape is displayed of the (first few)
eigenphase shifts for $L^{\pi }=0^{+}$ in $^{6}$Be in the full calculations,
i.e. with the maximal number of internal and asymptotic HH's. One notices
that the first $0^{+}$-resonance state of $^{6}$Be appears in the first
eigenchannel, and that a second (broad) resonance at a higher energy is
created in the second eigenchannel (see table \ref{tab:second_res} for
details on this resonance). The phase shifts in the higher eigenchannels
show a smooth behavior as a function of energy without a trace of resonances
in the energy range that we consider.

Table \ref{tab:Theory} compares the results of this work to those obtained
in other calculations, in particular by CSM \cite{kn:Csoto94}, by HHM \cite
{kn:Dani91} and by CCCM \cite{kn:CSM-tanaka}. In table \ref{tab:AM+exprm}
our results are compared with the experimental data available from \cite
{kn:Ajze88}. The agreement with the experimental energy and width of the
resonant states is reasonable. The difference between the experimental and
calculated energies of the 2$^{+}$-resonance states in both $^{6}$He and $%
^{6}$Be are probably due to the lack on LS-forces in the present
calculations.

It has been pointed out in \cite{kn:Dani91} as well as in \cite{kn:Csoto94},
that the barrier created by the three-cluster configuration is sufficiently
high and wide to accommodate two resonance states, the second one of which
is usually very broad. At first sight, this could be ascribed to an artifact
of the HH Model since large values of hypermomentum $K$ create a substantial
centrifugal barrier. However the CSM calculations \cite
{kn:CSM-Aoyama1,kn:CSM-Aoyama2}, which do not use HH's, also reveal such
resonances. Our calculations now also confirm the existence of a second very
broad resonance. A comparison of the resonance parameters with those of the
HH-method \cite{kn:Dani91} and of the CSM \cite{kn:Csoto94} is shown in
table \ref{tab:second_res}. The differences are most probably due to the
different descriptions of the system, as well as to the difference in
NN-forces.

\section{Conclusion}

In this work we have presented the results of the low-energy continuum
spectrum of $^{6}$He and $^{6}$Be as obtained in the three-cluster
description $\alpha +N+N$ \ within the Algebraic Model developed in \cite
{kn:ITP+RUCA1}.

It was shown that the range of the Pauli principle in three-cluster systems
extends well beyond the range in two-cluster models. This can be understood
by the fact that, even at high values of the hyperradius $\rho $, there is a
non-zero probability for one pair of clusters to be close to one another,
and thus within the range of the Pauli principle.

It was also shown that the folding approximation is already close to the
fully antisymmetrized potential for relatively compact three-cluster
configurations. Thus the folding approach, which is numerically much easier,
can approximate the exact potential with good precision in a large part of
the model space.

We have elaborated upon the role of the different HH states in the formation
and decay of resonance states in the three cluster continuum, and have
demonstrated that a sufficient number of HH's should be considered for a
proper description of the internal state. It was also seen that the number
of HH's in the asymptotic region is of lesser importance. It was
demonstrated that the HH basis thus represents an appropriate and convenient
tool for describing three-cluster systems, as it takes into account the
proper boundary conditions for the continuum of multi-cluster configurations.

The calculated values for energy and width of the resonance states agree
well with the observed experimental data. The results of the present model
are also in good accordance with those obtained in alternative models used
for the description of resonance states in six-nucleon systems.

The main conclusion to be drawn from this work is that the AM for
three-cluster configurations, which takes into account full
antisymmetrization between all particles, not only leads to meaningful
results, but provides a feasible alternative to other models.

\section{Acknowledgments}

This work was partially supported by an INTAS
Grant-(``INTAS-93-755-extension). One of the authors (V. V.) gratefully
acknowledges the University of Antwerp (RUCA) for a ``RAFO-gastprofessoraat
1998-1999'' and the kind hospitality of the members of the research group
``Computational Quantum Physics'' of the Department of ``Mathematics and
Computer Sciences'', University of Antwerp, RUCA, Belgium.

\begin{table}[tbp] \centering%
%
\caption{Enumeration of the number of Hyperspherical Harmonics $N_{h}$
for fixed value of $K$ and the accumulated number of
Hyperspherical Harmonics $N_{c}$.} 
\begin{tabular}{|c|c|c|c|c|c|c|c|}
\hline
$L^{\pi }=0^{+}$ & $K$ & 0 & 2 & 4 & 6 & 8 & 10 \\ \hline
& $N_{h}$ & 1 & 1 & 2 & 2 & 3 & 3 \\ 
& $N_{c}$ & 1 & 2 & 4 & 6 & 9 & 12 \\ \hline
$L^{\pi }=2^{+}$ & $K$ & 0 & 2 & 4 & 6 & 8 & 10 \\ \hline
& $N_{h}$ & - & 2 & 3 & 5 & 6 & 8 \\ 
& $N_{c}$ & - & 2 & 5 & 10 & 16 & 24 \\ \hline
\end{tabular}
\label{tab:HH state} 
\end{table}%
%

\begin{table}[tbp] \centering%
%
\caption{Matrix of effective charges for the 0$^{+}$-state in $^{6}Be$} 
\begin{tabular}{|c|c|c|c|c|}
\hline
$K;l_{1},l_{2}$ & $0;0,0$ & $2;0,0$ & $4;0,0$ & $4;2,2$ \\ \hline
$0;0,0$ & 7.274 & 0.006 & -0.129 & 1.414 \\ \hline
$2;0,0$ & 0.006 & 7.146 & -0.436 & 0.314 \\ \hline
$4;0,0$ & -0.129 & -0.436 & 7.428 & -0.877 \\ \hline
$4;2,2$ & 1.414 & 0.314 & -0.877 & 9.098 \\ \hline
\end{tabular}
\label{tab:table1} 
\end{table}%
%

\begin{table}[tbp] \centering%
%
\caption{Matrix of effective charges for the 2$^{+}$-state in $^{6}Be$} 
\begin{tabular}{|l|l|l|l|l|l|}
\hline
$K;l_{1},l_{2}$ & $2;2,0$ & $2;0,2$ & $4;2,0$ & $4;0,2$ & $4;2,2$ \\ \hline
$2;2,0$ & 7.253 & 0.400 & -0.224 & 0.546 & -0.751 \\ \hline
$2;0,2$ & 0.400 & 7.244 & -0.309 & -0.004 & -0.601 \\ \hline
$4;2,0$ & -0.224 & -0.309 & 6.942 & -0.186 & 0.431 \\ \hline
$4;0,2$ & 0.546 & -0.004 & -0.186 & 7.694 & -0.671 \\ \hline
$4;2,2$ & -0.751 & -0.601 & 0.431 & -0.671 & 7.345 \\ \hline
\end{tabular}
\label{tab:table2} 
\end{table}%
%

\begin{table}[tbp] \centering%
%
\begin{tabular}{|l|l|l|l|l|l|}
\hline
$K_{\max }^{(a)}$ & 0 & 2 & 4 & 6 & 8 \\ \hline
$E$, MeV & 1.434 & 1.314 & 1.304 & 1.298 & 1.292 \\ 
$\Gamma $, MeV & 0.075 & 0.082 & 0.084 & 0.085 & 0.087 \\ \hline
\end{tabular}
\caption{Parameters of the 0$^{+}$-resonance in $^6$Be as a function of
$K_{max}^{(a)}$ for fixed $K_{max}^{(i)}=8$}\label{tab:Conv_vs_asy_8}%
\end{table}%
%

\begin{table}[tbp] \centering%
%
\begin{tabular}{|c|c|c|c|c|c|c|}
\hline
$K_{\max }^{(a)}$ & 0 & 2 & 4 & 6 & 8 & 10 \\ \hline
$E$, MeV & 1.324 & 1.204 & 1.192 & 1.184 & 1.176 & 1.172 \\ 
$\Gamma $, MeV & 0.068 & 0.069 & 0.071 & 0.071 & 0.073 & 0.072 \\ \hline
\end{tabular}
\caption{Parameters of the 0$^{+}$-resonance in $^6$Be as a function of $K_{max}^{(a)}$ for fixed $K_{max}^{(i)}=10$}%
%
\label{tab:Conv_vs_asy_10}%
\end{table}%
%

\begin{table}[tbp] \centering%
%
\begin{tabular}{|c|c|c|c|c|c|c|}
\hline
$K_{\max }^{(i)}$ & 0 & 2 & 4 & 6 & 8 & 10 \\ \hline
$E$, MeV & - & 2.408 & 2.020 & 1.688 & 1.434 & 1.324 \\ 
$\Gamma $, MeV & - & 0.147 & 0.129 & 0.097 & 0.075 & 0.068 \\ \hline
\end{tabular}
\caption{Parameters of the 0$^{+}$-resonance in$^6$Be as a function of $K_{max}^{(i)}$ for $K_{max}^{(a)}=0$.}%
%
\label{tab:Conv_vs_int} 
\end{table}%
%

\begin{table}[tbp] \centering%
%

\begin{tabular}{|l|cc|cc|cc|}
\hline
\  & \multicolumn{2}{c|}{$^{6}$He; L$^{\pi }$ =2$^{+}$} & 
\multicolumn{2}{c|}{$^{6}$Be; L$^{\pi }$ =0$^{+}$} & \multicolumn{2}{c|}{$%
^{6}$Be; L$^{\pi }$ =2$^{+}$} \\ \hline
Method & $E$, MeV & $\Gamma$, MeV & $E$, MeV & $\Gamma$, MeV & $E$, MeV & $%
\Gamma$, MeV \\ \hline
AM & 1.490 & 0.168 & 1.172 & 0.072 & 3.100 & 0.798 \\ \hline
HHM\cite{kn:Dani91} & 0.75 & 0.04 &  &  &  &  \\ \hline
CSM \cite{kn:Csoto94} & 0.74 & 0.06 & 1.52 & 0.16 & 2.81 & 0.87 \\ \hline
CCCM\cite{kn:CSM-tanaka} & 0.73 & 0.07 &  &  &  &  \\ \hline
\end{tabular}
\caption{Energy and width of the resonace states in $^6$He and $^6$Be, obtained
by AM, HHM, CSM and CCCM}\label{tab:Theory} 
\end{table}%
%

\begin{table}[tbp] \centering%
%
\begin{tabular}{|c|cc|cc|}
\hline
\  & \multicolumn{2}{c|}{AM} & \multicolumn{2}{c|}{Experiment \cite
{kn:Ajze88}} \\ \hline
& $E$, MeV & $\Gamma $, MeV & $E$, MeV & $\Gamma $, MeV \\ \hline
$^{6}$He; L$^{\pi }$ =2$^{+}$ \  & 1.490 & 0.168 & 0.822$\pm $0.025 & 0.133$%
\pm $0.020 \\ \hline
$^{6}$Be; L$^{\pi }$ =0$^{+}$ & 1.172 & 0.072 & 1.371 & 0.092$\pm $0.006 \\ 
\hline
$^{6}$Be; L$^{\pi }$ =2$^{+}$ & 3.100 & 0.798 & 3.04$\pm $0.05 & 1.16$\pm $%
0.06 \\ \hline
\end{tabular}
\label{tab:AM+exprm} 
\caption{Energy and width of the resonant states in $^{6}$He and $^{6}$Be, obtained with AM
 and compared to experiment.} 
\end{table}%
%

\begin{table}[tbp] \centering%
%
\begin{tabular}{|l|l|cc|cc|cc|}
\hline
Nucleus & $L^{\pi }$ & \multicolumn{2}{c|}{AM} & \multicolumn{2}{l|}{HH\cite
{kn:Dani91}} & \multicolumn{2}{l|}{CSM\cite{kn:CSM-Aoyama1} \cite
{kn:CSM-Aoyama2}} \\ \hline
&  & $E$, MeV & $\ \Gamma $, MeV & $E$, MeV & $\Gamma $, MeV & $E$, MeV & $%
\Gamma $, MeV \\ \hline
$^{6}$He & 0$_{2}^{+}$ & 2.1 & 4.3 & 5.0 & 6.0 & 3.9 & 9.4 \\ \hline
$^{6}$He & 2$_{2}^{+}$ & 3.7 & 5.0 & 3.3 & 1.2 & 2.5 & 4.7 \\ \hline
$^{6}$Be & 0$_{2}^{+}$ & 3.5 & 6.1 &  &  &  &  \\ \hline
$^{6}$Be & 2$_{2}^{+}$ & 5.2 & 5.6 &  &  &  &  \\ \hline
\end{tabular}
\caption{Energy and width of  the second resonance states, obtained by AM, HHM and
CSM}\label{tab:second_res} 
\end{table}%
%

\begin{figure}[tbp]
\caption{Two configurations of the Jacobi coordinates for the three-cluster
system $\protect\alpha +N+N$.}
\label{fig:figure1}
\end{figure}

\begin{figure}[tbp]
\caption{Matrix elements $\left\langle n,(K,L=0)\left| \widehat{{\cal A}}%
\right| n,(K^{\prime },L=0)\right\rangle $ of the antisymmetrization
operator for the 0$^{+}$-state in $^{6}$He and $^{6}$Be}
\label{fig:figure2}
\end{figure}

\begin{figure}[tbp]
\caption{Matrix elements $\left\langle n,(K,\protect\alpha ,L=0)\left| 
\widehat{{\cal A}}\right| n,(K,\protect\alpha ,L=0)\right\rangle $ of the
antisymmetrization operator for the 0$^{+}$-state in $^{6}$He and $^{6}$Be}
\label{fig:figure3}
\end{figure}

\begin{figure}[tbp]
\caption{Comparison of matrix elements of the antisymmetrization operator
for the 0$^{+}$-state of $^{6}$He and $^{6}$Be obtained in the three-cluster
configuration $\protect\alpha +N+N$ (for $K=0$) and the two-cluster
configuration $\protect\alpha +2N$}
\label{fig:figure4}
\end{figure}

\begin{figure}[tbp]
\caption{Diagonal matrix elements of the potential energy operator for the 0$%
^{+}$-state of $^{6}$Be and $^{6}$He (for $K=0$) with full
antisymmetrization compared to those within the folding model}
\label{fig:figure5}
\end{figure}

\begin{figure}[tbp]
\caption{Off-diagonal matrix elements of the potential energy operator for
the 0$^{+}$-state of $^{6}$Be and $^{6}$He (for $K=0$) with full
antisymmetrization compared to those within the folding model}
\label{fig:figure6}
\end{figure}

\begin{figure}[tbp]
\caption{Matrix elements of the potential energy operator for the 0$^{+}$%
-state of $^{6}$Be and $^{6}$He, diagonal in $n$, with full
antisymmetrization}
\label{fig:figure7}
\end{figure}

\begin{figure}[tbp]
\caption{Contributions in the folding model of the main components to the
diagonal matrix elements of the potential energy operator for the 0$^{+}$%
-state of $^{6}$Be and $^{6}$He (for $K=0$) }
\label{fig:figure8}
\end{figure}

\begin{figure}[tbp]
\caption{Eigenphase shifts for the $0^{+}$-state of $^{6}$Be for different
values of $K_{max}^{(a)}$}
\label{fig:figure9}
\end{figure}

\begin{figure}[tbp]
\caption{The\ eigenphases for the 0$^{+}$-state in $^{6}$Be.}
\label{fig:figure10}
\end{figure}

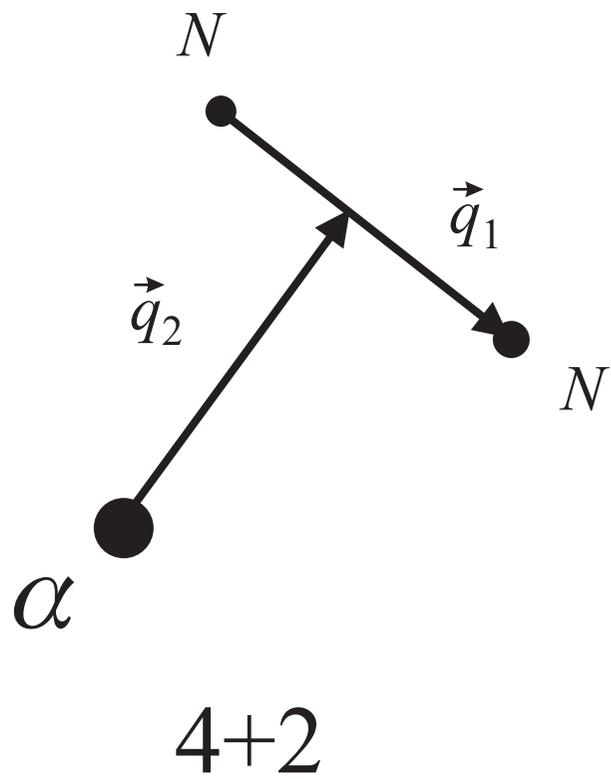
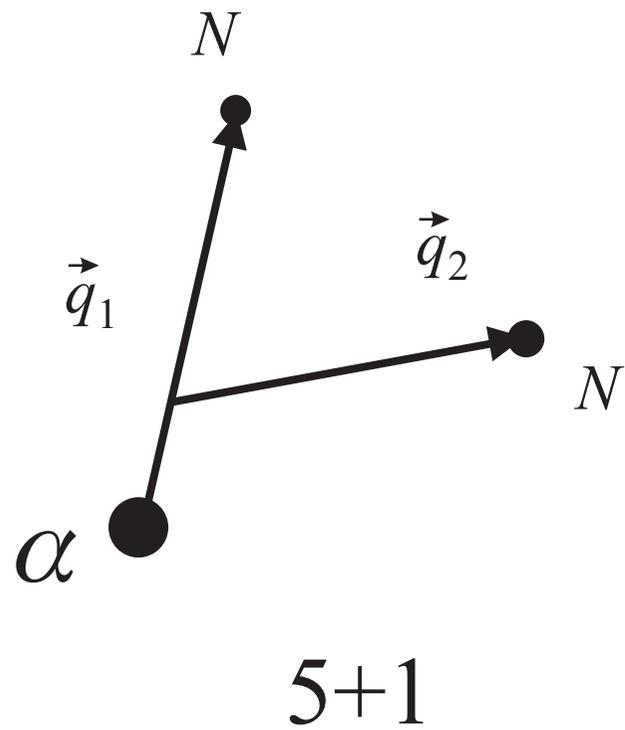
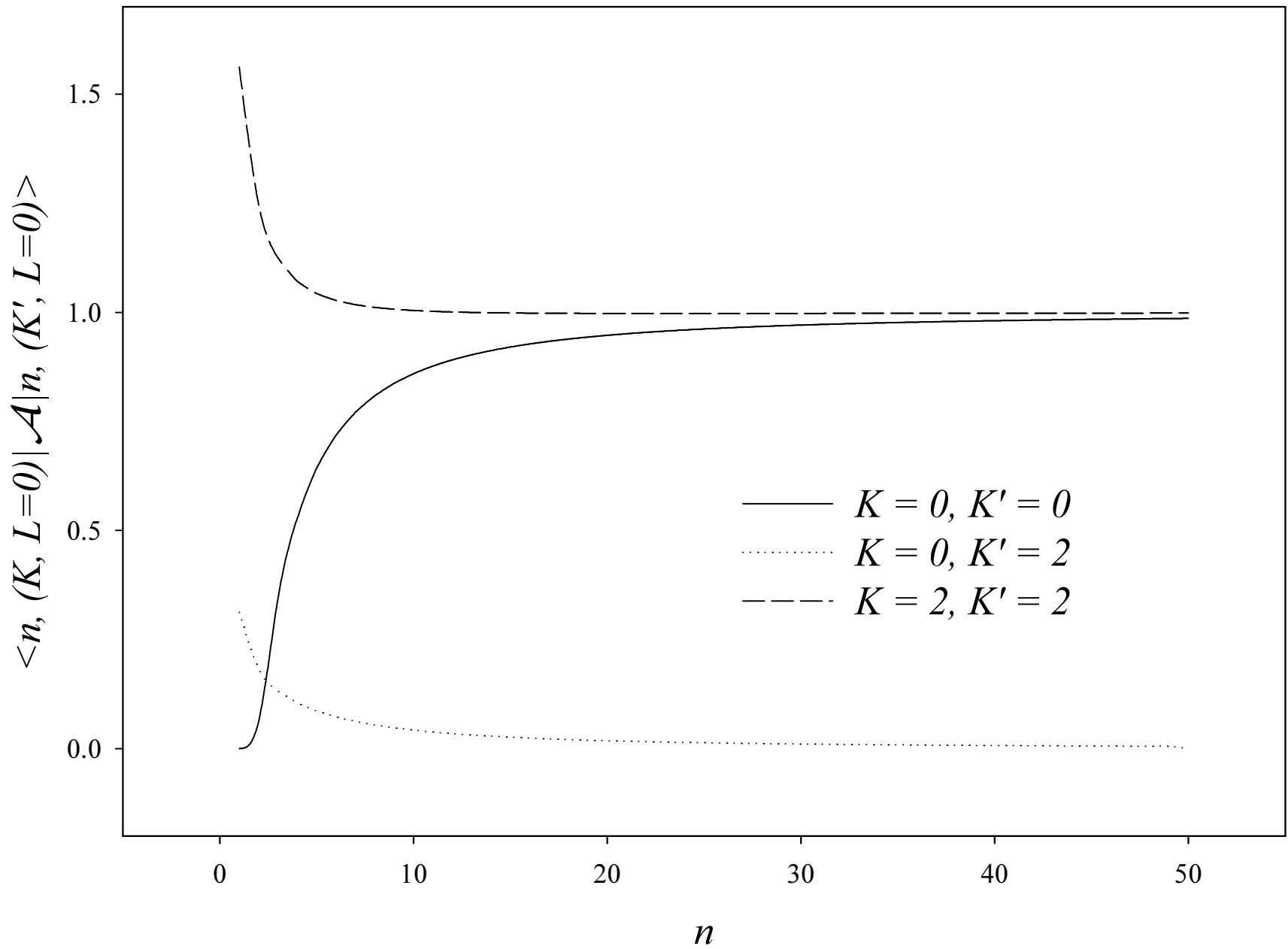
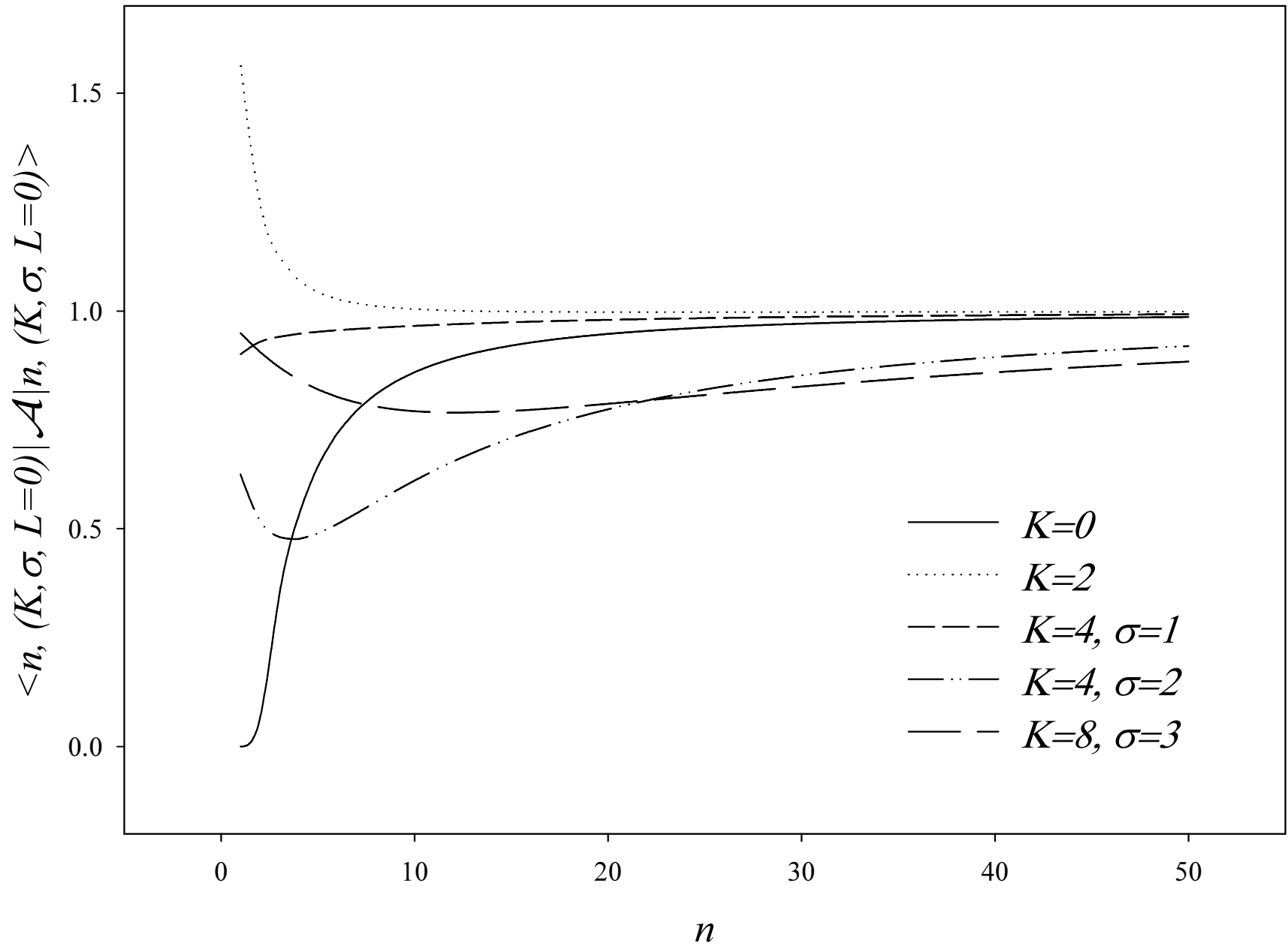
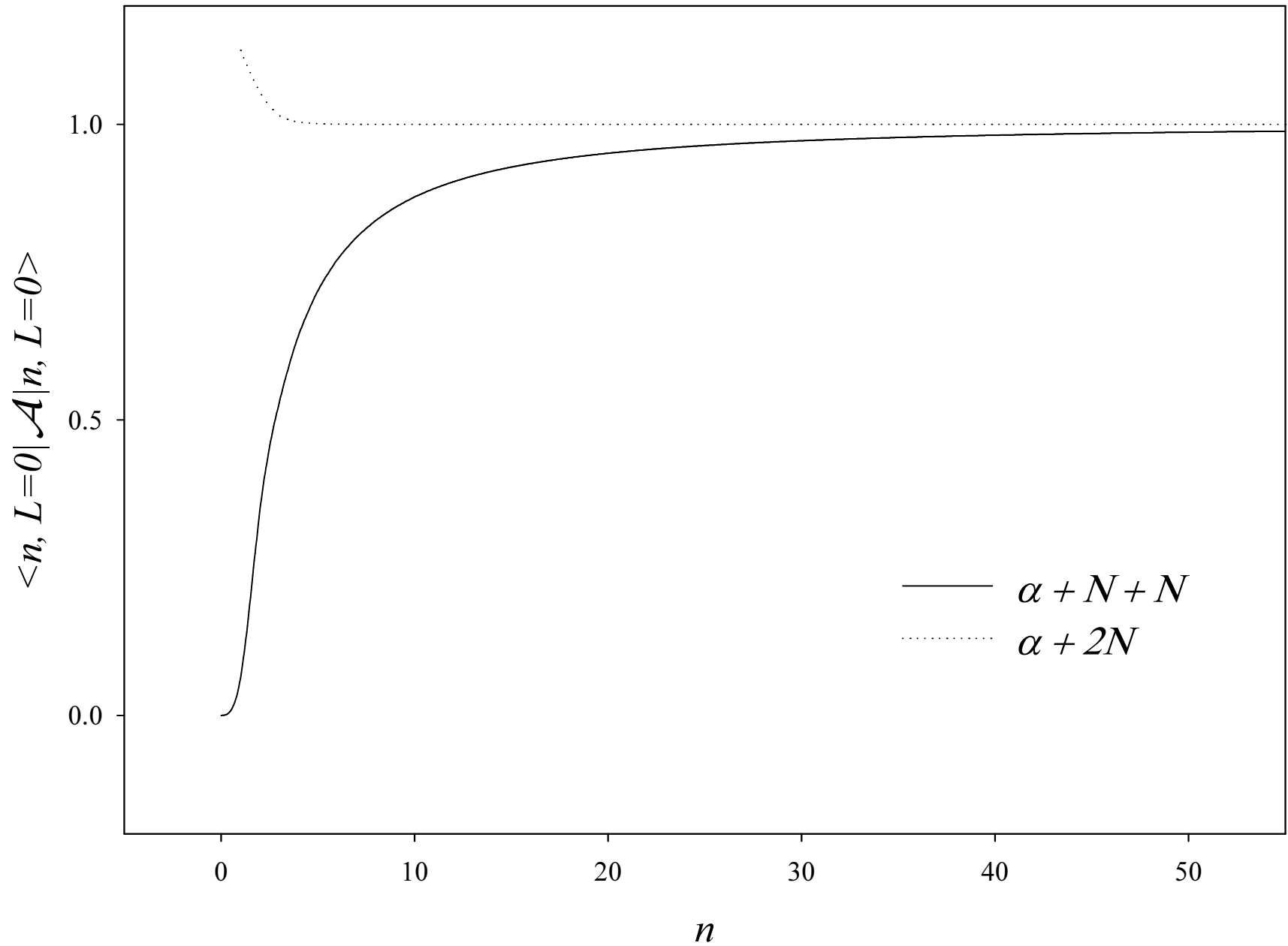
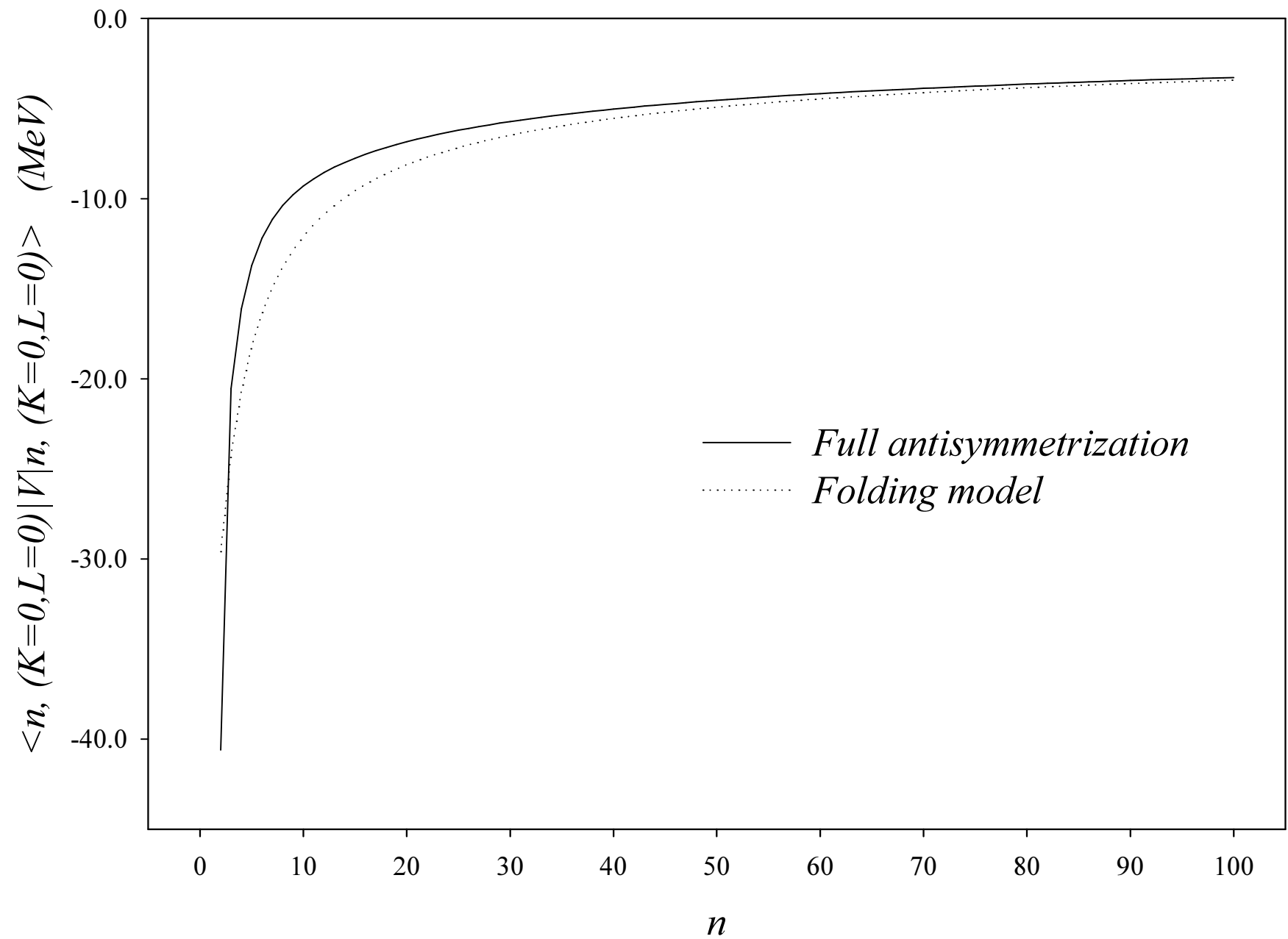
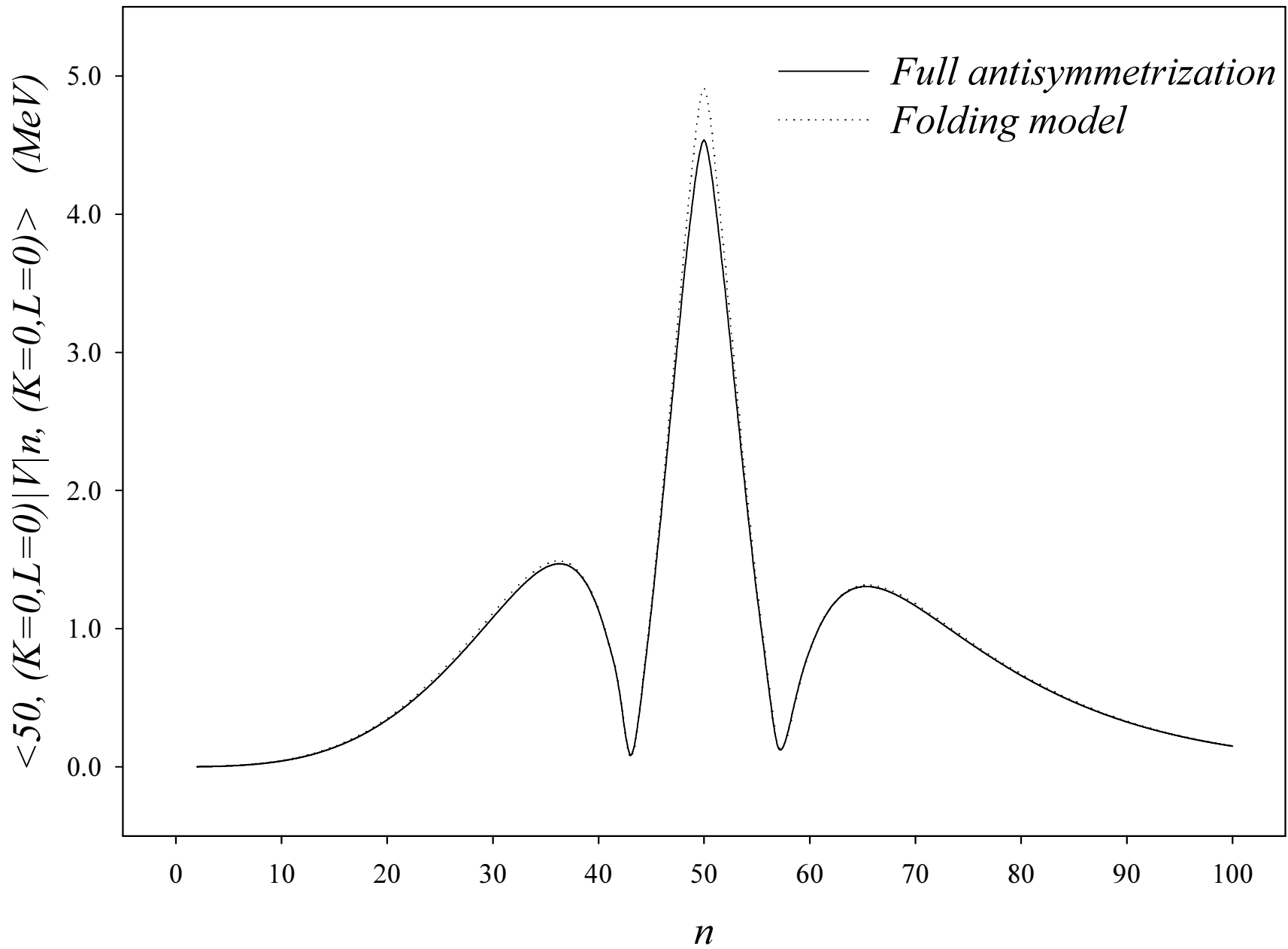
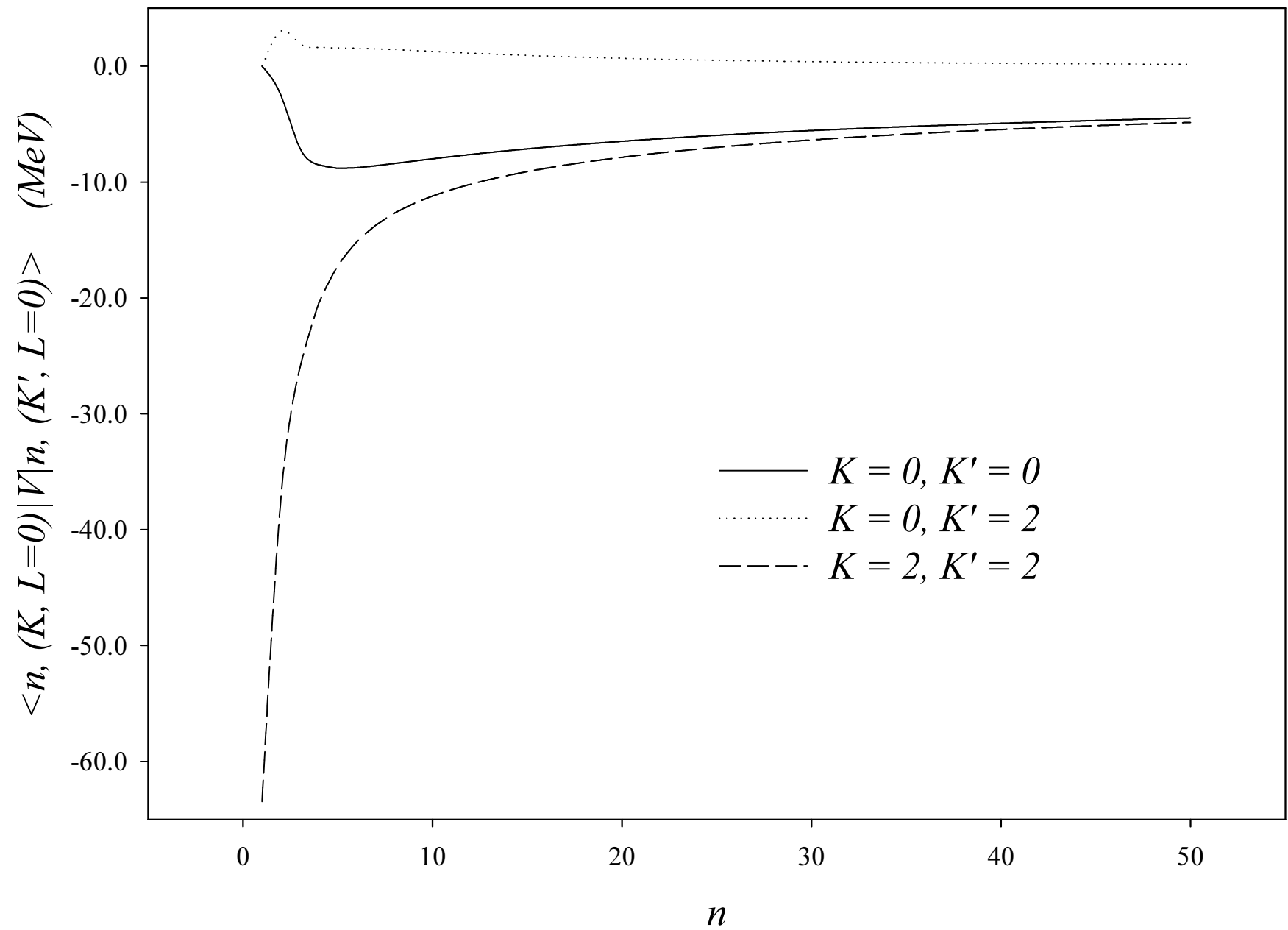
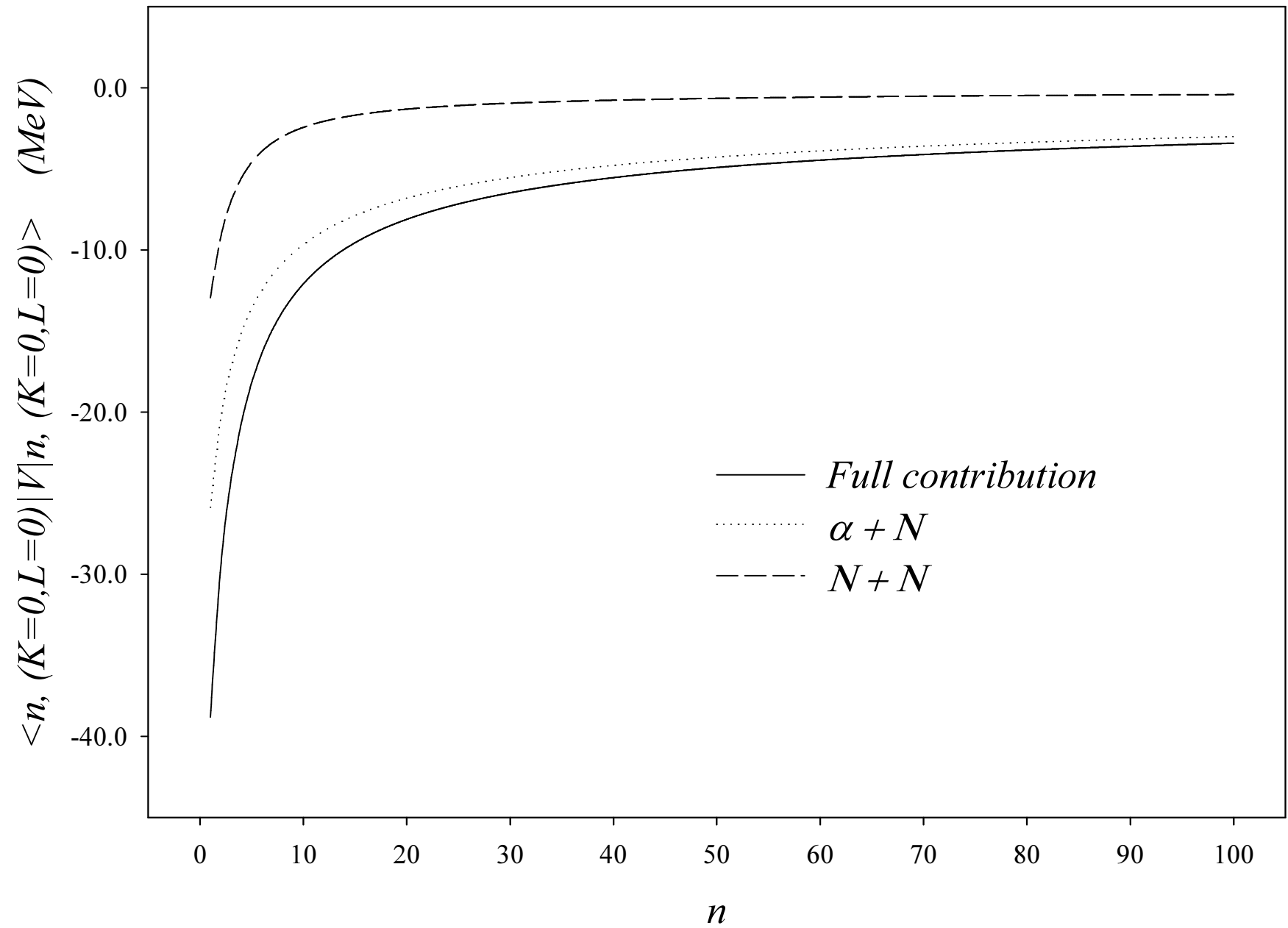
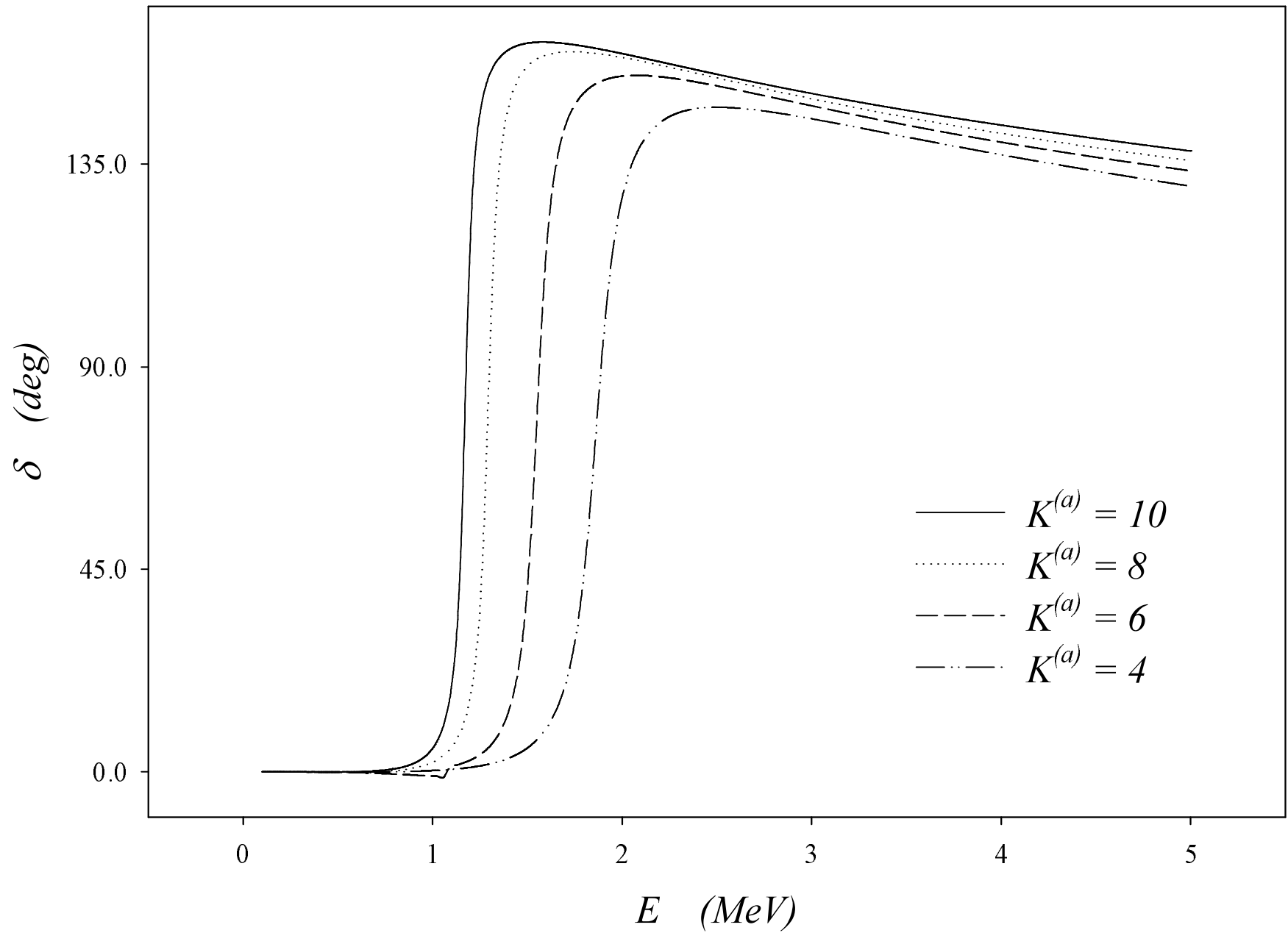
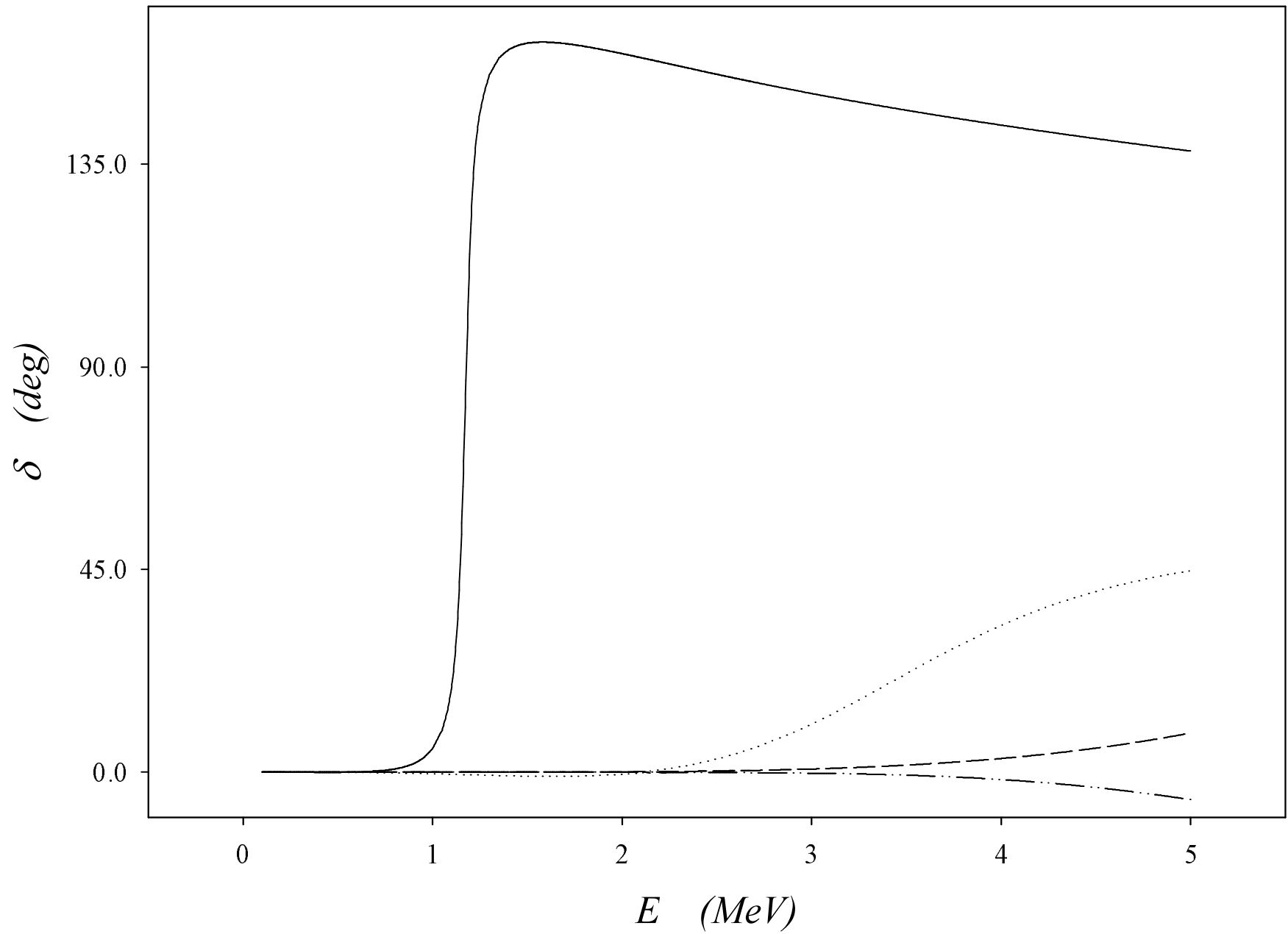
\end{document}